# ON DIMENSIONAL REGULARIZATION AND MATHEMATICAL RIGOUR

F.V. Tkachov[#1]

*Institute for Nuclear Research of the Russian Academy of Sciences, Moscow 117312, Russia*

The controversy concerning the phenomenon of breakdown of dimensional regularization in the problems involving asymptotic expansions of Feynman diagrams in non-Euclidean regimes is discussed with some pertinent bibliographic comments.

I would like to clarify some points in the discussion of whether or not the dimensional regularization [1] (see also [2]) breaks down in the explicit and very simple example presented in [3]. The example is a one-loop scalar triangle diagram with only one non-zero internal mass $m$ in the horizontal "gluon" propagator in the figure below. One considers the case when $Q^2 = -(p_1 - p_2)^2 \to \pm\infty$ with $m$ fixed and $p_i^2 = 0$, and one attempts to construct a perfectly factorized asymptotic expansion (a one-loop Sudakov analogue of OPE). Then the soft singularity corresponding to the gluon can not be regulated by dimensional continuation after the first subtraction (corresponding to collinear singularities) is done.

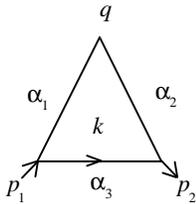

Surprisingly, our finding was disputed in [4] which concluded that "one should not be afraid of using dimensional regularization in practical calculations connected with the Sudakov limit".

Referring the reader for a detailed critique of [4] to [5], I'd like to discuss some additional points.

Without further ado, let me present an exact expression (pointed out by N. Ussyukina) for the triangle graph in terms of Feynman parameters in which integrations over two of the three Feynman parameters have been done (an interested reader will easily reproduce it):

$$\frac{1}{Q^{2(1+\varepsilon)}} \int_0^1 \frac{d\alpha}{\alpha - \lambda} \times (1-\alpha)^{-\varepsilon} \frac{[\alpha^{-\varepsilon} - \lambda^{-\varepsilon}]}{\varepsilon}, \qquad (1)$$

where $\lambda = m^2 / Q^2 > 0$, and a trivial overall coefficient is dropped. The representation of the diagram in terms of hypergeometric functions quoted in [4] can be directly obtained from this.

If one attempts to extract the powers and logarithms of $\lambda \to 0$ prior to integration over $\alpha$, then one encounters the singularity at $\alpha = 0$ generated after the denominator is Taylor-expanded in $\lambda$, and that singularity is not regulated by $\varepsilon \neq 0$. The simplest manifestation of this is that one will not be able to perform the integration in the expanded expression termwise—even at $\varepsilon = \frac{1}{2}(4-D) \neq 0$— without running into infinities. It is hard to imagine how the standpoint of ref. [4] could be defended in view of the explicit evidence provided by eq. (1).

To reiterate: if one attempts to expand the above triangle graph in powers and logarithms of $\lambda$ (i.e. to perform a *factorization*), then the first subtraction one has to make corresponds to the collinear singularities, and after collinear subtractions, the resulting expression—as was found in [3]—contains a singularity that is not regulated by dimensional continuation. This pathology persists (as it should) irrespective of the representation one uses (e.g. it is manifest in the above integral representation). ∎

It should be stressed that the errors in [4] were committed at the level of interpretation rather than calculations. Indeed, ref. [4] failed to point out any concrete mistakes in [3], and the simple calculation performed in [4] (which correctly found the presence of the second logarithm in the expansion—a logarithm that is clearly not associated with poles in $\varepsilon$) confirms rather than disproves the conclusions of [3].

Ref. [4] contains statements that are breathtaking to an expert—e.g. that some UV divergences may cancel IR ones (recall the proverbial apples and oranges)—a revelation declared in [4] to constitute a "principal new feature" of the expansion problem in non-Euclidean asymptotic regimes (see [5] for an explanation of the origin of this fallacy).

The misinterpretations of [4] are not accidental. They are fortified by references to the book [6] that purports to provide an uncompromisingly "rigorous" account of the state of the art in the theory of multiloop calculations and asymptotic properties of Feynman diagrams. Because all our

---
[#1] E-mail: ftkachov@ms2.inr.ac.ru

knowledge comes at second hand in some way, a number of theorists took for granted the awesome rigor of [6] and were misled by [4].

Therefore, I would like to take the bull by the horns and examine the book [6] itself. A book review being a perfectly legitimate scientific genre, it is also clear that (i) to trace the origins of a misconception is no less instructive than those of a discovery; (ii) the apparent discrepancy between the grandiose formalism of [6] and the elementary nature of the errors of [4] may be the most puzzling aspect of the current dispute; (iii) references to the "rigor" of [6] constitute to many (let us face it) the single most convincing argument of [4].

The proof of the pudding is in the eating. Why mathematical research into the nature of multiloop Feynman diagrams should be judged otherwise? Indeed, it is useful inasmuch as one can learn something from it about how to deal with those notoriously complex beasts. (One could argue that such a research could be interesting from the point of view of pure mathematics, but there is nothing in [6] worth mentioning in that respect.)

The design of ref. [6] is simple. It outlines a method of purportedly rigorous definition of dimensionally regulated Feynman diagrams, and then it sketches how the method *could* be used to "rigorously prove" many well-known results in applied QFT. An unwary reader will be impressed by the formidable formalism, while the bibliographic comments are phrased in such a way as to leave no doubt that the formulae calculationists had been using for years in highly non-trivial calculations, did not, really, exist before they were declared "rigorously proved" by the author of [6].

A closer examination, however, reveals the following interesting circumstance. *All* more or less significant results described in [6] (starting from the results on singular structure of Feynman diagrams [8]) were found by other authors by other methods (which the author of [6] did not bother to describe), in some cases confirmed by as many as three groups (e.g. [7], [9], [10]); have been checked for self-consistency in dozens of sophisticated calculations (e.g. [12]); the complexity of those results should leave no doubt that their authors certainly knew what they were doing. To translate their heuristics (if one may thus qualify their exact algorithms) into a "rigorous" lower level formalism is only a matter of having spare time and nothing better (sic!) to do.

Further, consider the method of defining dimensionally regulated diagrams via the $\alpha$-parametric representation described in [6]. It implements, in a most straightforward fashion, the elementary idea of splitting the integration domain into as many sectors à la Sudakov-Hepp as needed to make the integral in each subdomain absolutely convergent for some values of $D$, and then using trivial tricks (integration by parts) to perform analytical continuation. (Apart from minor modifications, the formalism of this kind was thoroughly worked through in [11].) This is extremely cumbersome (like programming in machine codes)—and it is also obvious that an intermediate layer of abstraction is missing to hide the ugly details (cf. high level programming languages). The method of [6] is also totally impractical—the resulting formalism is far too cumbersome to induce any calculationist to use it to check a new trick. The reader is left in disappointment to decide whether or not to trust the outlines of proofs presented in [6].

An examination of the journal publications on which ref. [6] is based reveals that the actual proofs are as a rule omitted there as being "technical" while most of the text consists of lengthy formulations of "theorems" that simply rephrase other people's results in an obscure formal manner. But what is then the claim of the author of [6]? If the proofs are not worth publishing in full, why are they as important as ref. [6] attempts to convince us?

And the crucial question: if those proofs are so cumbersome that cannot be presented in full, who and how should prove that they are correct?

An experienced theorist would of course never rely on such an evidence and would always prefer concrete checks of his/her findings (calculations, formulas etc.) by another method. Meaningful—even if not completely formalized—heuristics are incomparably more useful in that respect than the sterile rigor of obscure "proofs".

Ref. [6] also contains a compilation of formulas for multiloop calculations. The compilation is, understandably, rather indiscriminate—it takes a first-hand experience of real-life calculations to make such a compilation useful. Instead, the emphasis of [6] is on justification of the tricks within dimensional regularization. However, the "justification" is mostly reduced to formal references to the cumbersome constructions of dimensionally regularized diagrams described previously.

The last chapters of [6] present a version of "proof" of general Euclidean asymptotic expansions of Feynman diagrams in the MS scheme discovered and verified in different ways by three groups of authors [7], [9], [10], where all the formulas that are needed in practical calculations were found and subsequently verified in several record-setting 3- to 5-loop calculations (for a review see [7]). Ref. [6] treats the pioneer publications—the publications that contributed to defining the state of the art in multiloop calculations—with dismissive (sometimes grossly misleading) small-print comments. Apparently, those papers are not up to the standards of "rigor" of [6]. But a more likely reason is that the book [6] contains hardly any genuinely new ideas or results as compared with the original publications of the other authors. In fact, the version of proofs described in [6] is extremely cumbersome and exhibits the worst features of the BPHZ method (cf. the discussion in [7])—for lack of space I cannot quote e.g. the incomprehensible definition of IR subgraphs given in [6], to be compared with the simple criterion of the original publications (see [7]). And again, neither ref. [6] nor the journal publications it is based upon contain complete proofs.

The nature of the book [6] explains the fiasco of [4]: all of the meaningful problems considered and "proved" in [6] had already been posed, thought out and solved by others. Ref. [4] marked the first time the author of [6] ventured to consider a genuinely novel and delicate problem.

Mathematics is not about rigor—or, at least, not about that bad kind of rigor that hides a lack of genuine ideas. Like any natural science, a good mathematics is first and foremost about understanding, about solving difficult problems, and about making discoveries.

Unfortunately, a student of Applied Quantum Field Theory seeking insight into the nature of multiloop Feynman diagrams, will not profit from [6]. *Motivation*, *heuristics*, and *understanding* are the concepts banished from the compilation [6] with the utmost "rigor" of an empty formalism.

*Acknowledgments*. I thank E. Kazes for lending me a copy of ref. [6], J. Collins for discussions, and N. Ussyukina for pointing out the representation (1).

> for there must be also heresies among you, that they which are approved may be made manifest among you
> 1st Corinthians, 11:19